\documentclass[12pt]{article}

\newif\ifieeetrans
\ieeetransfalse 

\newif\ifllncs
\llncsfalse  

\ifllncs\fi
\ifieeetrans\fi

\ifieeetrans
\ifllncs
\bye                            
\fi
\fi

\newif\ifart
\artfalse
\ifieeetrans
\else
\ifllncs
\else
\arttrue
\fi
\fi

\ifieeetrans
\newcommand{\grphext}{_ieeetran}
\else
\ifllncs
\newcommand{\grphext}{_llncs}
\else
\newcommand{\grphext}{}
\fi
\fi

\ifieeetrans

\IEEEoverridecommandlockouts

\fi

\newif\ifpriorities
\prioritiestrue 

\usepackage{graphics}
\usepackage[matrix,arrow,curve]{xy}
\usepackage{amssymb}
\usepackage{amsmath}
\ifllncs
\usepackage{wrapfig}
\else
\usepackage{amsthm}
\fi
\usepackage{url}
\ifieeetrans
\else
\usepackage{smallsubsub}
\fi
\usepackage{hyperref}\hypersetup{colorlinks=true,}

\newtheorem{thm}{Theorem}
\newtheorem{lem}[thm]{Lemma}
\ifieeetrans
\newcommand{\para}[1]{\emph{#1}}
\else
\newcommand{\para}[1]{\paragraph{#1}}
\fi
\newcommand{\cpsa}{\textsc{cpsa}}
\newcommand{\pvs}{\textsc{pvs}}
\newcommand{\cn}[1]{\ensuremath{\operatorname{\mathsf{#1}}}}
\newcommand{\fn}[1]{\ensuremath{\operatorname{\mathit{#1}}}}
\newcommand{\srt}[1]{\ensuremath{\mathsf{#1}}}
\newcommand{\typ}{\mathbin:}

\newcommand{\enc}[2]{\ensuremath{\{\!|#1|\!\}_{#2}}}
\newcommand{\invk}[1]{{#1}^{-1}}
\newcommand{\tg}[1]{\cn{g}_{#1}}
\newcommand{\inbnd}{\mathord -}
\newcommand{\outbnd}{\mathord +}
\newcommand{\nat}{\ensuremath{\mathbb{N}}}

\newcommand{\all}[1]{\forall#1\mathpunct.}
\newcommand{\some}[1]{\exists#1\mathpunct.}

\newcommand{\up}{\mathord\uparrow}

\newcommand{\thy}[1]{\ensuremath{T_\mathit{#1}}}
\newcommand{\tbnd}{\thy{bnd}}
\newcommand{\tstate}{\thy{state}}
\newcommand{\tannot}{\thy{annot}}

\newcommand{\att}{\cn{att}}
\newcommand{\key}{\srt{A}|\srt{S}}
\newcommand{\base}{\key|\srt{D}|\srt{E}}
\newcommand{\boot}{\cn{bt}}
\newcommand{\extend}{\cn{ex}}
\newcommand{\tran}{\ensuremath{\leadsto}}
\newcommand{\pth}{\ensuremath{\pi}}
\newcommand{\pcr}{\fn{pcr}}
\newcommand{\evt}{\fn{evt}}
\newcommand{\sdom}{\fn{Dom}}
\newcommand{\ra}{\fn{ra}}

\newcommand{\skel}{\ensuremath{\mathbb{A}}}

\newcommand{\skelB}{\ensuremath{\mathbb{B}}}

\newtheorem{goal}{Security Goal}

\ifpriorities
\newcommand{\initstate}{p}
\else
\newcommand{\initstate}{\boot}
\fi

\title{A Hybrid Analysis for \\ Security Protocols with
  State\thanks{This work partially supported by the US National
    Security Agency, and partially supported by the National Science
    Foundation under grant CNS-1116557.  Authors' email addresses:
    \{guttman,prowe,ramsdell\}@mitre.org, \{dd,guttman\}@wpi.edu.}}
\author{John D.~Ramsdell \and Daniel J.~Dougherty \and \ifllncs\\\fi
  Joshua D.~Guttman \and Paul D.~Rowe}

\ifllncs
\institute{The MITRE Corporation and Worcester Polytechnic Institute}
\fi

\begin{document}
\maketitle
\begin{abstract}  \ifllncs\sloppypar\fi
  Cryptographic protocols rely on message-passing to coordinate
  activity among principals.  Many richly developed tools, based on
  well-understood foundations, are available for the design and
  analysis of pure message-passing protocols.  However, in many
  protocols, a principal uses non-local, mutable state to coordinate
  its local sessions.  Cross-session state poses difficulties for
  protocol analysis tools.

  We provide a framework for modeling stateful protocols, and a hybrid
  analysis method.  We leverage theorem-proving---specifically,
  PVS---for reasoning about computations over state.  An
  ``enrich-by-need'' approach---embodied by CPSA---focuses on the
  message-passing part.  The Envelope Protocol, due to Mark Ryan
  furnishes a case study.

\end{abstract}

\label{sec:intro}

\noindent Protocol analysis is largely about message-passing in a
model in which every message transmitted is made available to the
adversary.  The adversary can deliver the messages transmitted by the
regular (i.e.~compliant) principals, if desired, or not.  The
adversary can also retain them indefinitely, so that in the future he
can deliver them, or messages built from them, repeatedly.

However, some protocols also interact with long-term state.  For
instance, the Automated Teller Machine protocols interact with the
long-term state stored in banks' account databases.  Protocol actions
are constrained by that long-term state; for instance, an ATM machine
will be told not to dispense cash to a customer whose account has
insufficient funds.  Protocol actions cause updates to long-term
state; for instance, a successful withdrawal reduces the funds in the
customer's account.  State-manipulating protocols are important to
electronic finance and commerce.  They are also important in trusted
computing, i.e.~in systems using Trusted Platform Modules for
attestation and secrecy.  Indeed, as software interacts with
real-world resources in interoperable ways, cryptographic protocols
that manipulate long-term state will be increasingly central.

Long-term state is fundamentally different from message passing.  The
adversary can always choose to redeliver an old message.  But he
cannot choose to redeliver an old state; for instance, the adversary
in an ATM network cannot choose to replay a withdrawal, applying it to
a state in which he has sufficient funds, in case he no longer does.
Regular principals maintain long-term state across protocol executions
in order to constrain subsequent executions, and ensure that future
runs will behave differently from past runs.

The Cryptographic Protocol Shapes Analyzer~\cite{cpsa09} ({\cpsa}) is
our program for automatically characterizing the possible executions
of a protocol compatible with a specified partial execution.  It is
grounded in strand space theory.  There exists a mathematically
rigorous theory~\cite{cpsatheory11} that backs up the implementation
of {\cpsa} in Haskell, and proves the algorithm produces
characterizations that are complete, and that the algorithm enumerates
these characterizations.

Part of state manipulation can be encoded by message-passing.  In this
``state-passing style,'' reception of a message bearing the state
represents reading from the state, and transmission of an updated
state as a message represents writing to the state.  These conventions
help {\cpsa} analyze protocols with state.  If a protocol interacts
with the state, we add state-bearing receive/transmit event pairs to
its roles, and {\cpsa} attempts to find paths through state space as
it generates executions.  However, {\cpsa} constructs some executions
which are in fact not possible.  In these executions, a state-bearing
message is transmitted from one node and then received by two
different state-receiving nodes.

{\cpsa} does not recognize that this is not possible in a
state-history, and thus provides only an approximate analysis.
Showing the correctness of the protocol requires a more refined
analysis.

\para{Our contribution.}
We apply {\cpsa} to a system that relies on state, coupling {\cpsa}
with the Prototype Verification System~\cite{cade92-pvs} ({\pvs})
proof assistant.

We specified a version of strand space theory in {\pvs}.  On top of
this theory, we encoded the result of a {\cpsa} analysis run as a
formula in the {\pvs} logic.  This formula is justified by the {\cpsa}
completeness result~\cite{Ramsdell13}.  We then use this formula as an
axiom in {\pvs}.  Proofs using this axiom may imply the existence of
additional message transmission/receptions, leading to an enriched
{\cpsa} analysis.  In this way the theorem-proving and
execution-finding analysis activities cooperate, over the common
semantic foundation of strand space theory.  Hence, the combination is
semantically sound.

%

\para{Outline of the Analysis.}
Our paradigm is \cpsa's enrich-by-need approach~\cite{Guttman12a}.
\ifllncs
\begin{wrapfigure}{l}{.5\textwidth}\small\vspace{-6mm}
\[
\xymatrix@R=1mm@C=1mm{
  \tbnd\ar[rr]\ar[rd]&&\tbnd(\Pi)\ar[rd]\\
  &\tannot\ar[rr]&&\tannot(\Pi,\leadsto)\\
  \tstate\ar[rr]\ar[ru]&&\tstate(\leadsto)\ar[ru]}
\]
\caption{Theory Inclusions}\label{fig:theory refinements}
\vspace{-7mm}
\end{wrapfigure}
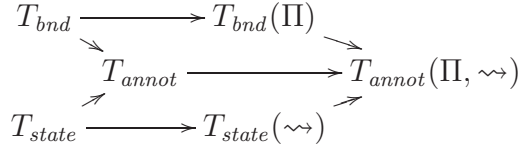
\fi
That is, we ask:  What kinds of executions are possible, assuming that
a particular pattern of events has occurred?  To verify authentication
properties, we observe that all executions contain certain required
events.  To verify confidentiality properties, we consider patterns
that include a disclosure, and observing that no executions are
possible.  Our method involves a conversation (so to speak) between
{\cpsa} and {\pvs}.  The main steps are:
\begin{enumerate}
  \item Within \pvs\ we define theories (i) $\tbnd$ of strand spaces
  and protocol executions (``bundles'') and (ii) $\tstate$ of
  transition relations and their state histories (see
  Fig.~\ref{fig:theory refinements}).  $\tannot$ is their union, a
  theory of protocol executions where some protocol steps are
  annotated with a state transition.  Augmenting $\tbnd$ with
  information about a protocol $\Pi$ produces $\tbnd(\Pi)$.
  Augmenting $\tstate$ with information about a particular transition
  relation $\leadsto$ produces $\tstate(\leadsto)$.  The union of
  $\tannot$, $\tbnd(\Pi)$, and $\tstate(\leadsto)$ is
  $\tannot(\Pi,\leadsto)$.

  Our {\pvs} theories are in fact somewhat coarser than this.
  \item Within the state transition theory $\tstate(\leadsto)$, we
  prove lemmas in {\pvs} such as Lemma~\ref{lem:prefix boot extend}
  below.  Some of their consequences in the annotated protocol theory
  $\tannot(\Pi,\leadsto)$ use only the limited vocabulary of
  $\tbnd(\Pi)$; we call them \emph{bridge lemmas}.
  Lemma~\ref{lem:bridge} is a bridge lemma.  They bring information
  back from the state world to the protocol world.
  \item Independently, {\cpsa} analyzes the protocols, with
  state-manipulation modeled as message-passing, but without any
  special knowledge about state transition histories.  A sentence,
  called a \emph{shape analysis
    sentence}~\cite{Ramsdell12,Guttman12a}, summarizes its results in
  a sentence in the language of $\tbnd(\Pi)$.  A shape analysis
  sentence, such as Lemma~\ref{lem:alice shape}, is used as an axiom
  in proofs within {\pvs}.
  \item Using bridge lemmas and state analysis sentences jointly, we
  infer conclusions about protocol runs in $\tbnd(\Pi)$.  If we prove
  a contradiction, that shows that the situation given to {\cpsa}
  cannot in fact occur.  Otherwise, we may prove that additional
  message transmissions and receptions occurred, as in
  Thm.~\ref{thm:inferred strand}.
  \item We incorporate these additional nodes into a new {\cpsa}
  starting point, and allow {\cpsa} to draw conclusions.  Additional
  round trips are possible.
\end{enumerate}
\ifllncs
\else
\begin{figure}[t]
\[
\xymatrix@R=1mm@C=1mm{
  \tbnd\ar[rr]\ar[rd]&&\tbnd(\Pi)\ar[rd]\\
  &\tannot\ar[rr]&&\tannot(\Pi,\leadsto)\\
  \tstate\ar[rr]\ar[ru]&&\tstate(\leadsto)\ar[ru]}
\]
\caption{Theory Inclusions}\label{fig:theory refinements}
\vspace{-7mm}
\end{figure}
\fi

\ifllncs
\else
\para{Structure.}
The body of this paper describes an
application of our method to the Envelope Protocol, a protocol that
interacts with a Trusted Platform Module (TPM) to achieve an important
security goal.  Section~\ref{sec:envelope} describes the
protocol~$\Pi$.  Section~\ref{sec:tpm} describes our TPM model,
$\tstate(\leadsto)$.  Section~\ref{sec:strand spaces}
presents the theory of bundles $\tbnd$ encoded within
{\pvs}, and specializes this to $\tbnd(\Pi)$, demonstrating
our main trick of including state-bearing receive-transmit pairs to
encode the state transitions.  Section~\ref{sec:cpsa} describes
{\cpsa}, our protocol analysis tool and what results {\cpsa} infers in
$\tbnd(\Pi)$. Section~\ref{sec:pvs} links the state world
and the protocol world $\tannot(\Pi,\leadsto)$.  The
relevant bridge lemma is stated and applied to prove the Envelope
Protocol security goal.
\fi

\section{The Envelope Protocol}
\label{sec:envelope}

The proof of an important security goal of the Envelope
Protocol~\cite{ables2010escrowed} was the focus of most of our effort.
The protocol allows someone to package a secret such that another
party can either reveal the secret or prove the secret never was and
never will be revealed.

\para{Protocol motivation.}  The plight of a teenager motivates
the protocol.  The teenager is going out for the night, and her
parents want to know her destination in case of emergency.  Chafing at
the loss of privacy, she agrees to the following protocol.  Before
leaving for the night, she writes her destination on a piece of paper
and seals the note in an envelope.  Upon her return, the parents can
prove the secret was never revealed by returning the envelope
unopened. Alternatively, they can open the envelope to learn her
destination.

The parents would like to learn their daughter's destination while
still pretending that they have respected her privacy. The parents are
thus the adversary.  The goal of the protocol is to prevent this
deception.

\para{Necessity of long-term state.}  The long-term state is the
envelope.  Once the envelope is torn open, the adversary no longer has
access to a state in which the envelope is intact.  A protocol based
only on message passing is insufficient, because the ability of the
adversary monotonically increases. At the beginning of the protocol
the adversary can either return the envelope or tear it. In a purely
message-based protocol the adversary will never lose these abilities.

\para{Cryptographic version.}  The cryptographic version of this
protocol uses a TPM to achieve the security goal.  Here
we restrict our attention to a subset of the TPM's functionality. In
particular we model the TPM as having a state consisting of a single
Platform Configuration Register (PCR) and only responding to five
commands.  A \texttt{boot} command sets the PCR to a known
value.  The \texttt{extend} command takes a piece of data, $d$, and
replaces the current value $\fn{val}$ of the PCR with the hash of $d$
and $\fn{val}$, i.e. $\#(d,\fn{val})$.  In fact, the form of
\texttt{extend} that we model, which is an \texttt{extend} within an
encrypted session, also protects against replay.  These are the only
commands that alter the value in a PCR.

The TPM provides other services that do not alter the PCR.  The
\texttt{quote} command reports the value contained in the PCR and is
signed in a way as to ensure its authenticity.  The \texttt{create
  key} command causes the TPM can create an asymmetric key pair where
the private part remains shielded within the TPM.  However, it can
only be used for decryption when the PCR has a specific value.  The
\texttt{decrypt} command causes the TPM to decrypt a message using
this shielded private key, but only if the value in the PCR matches
the constraint of the decryption key.

In what follows, Alice plays the role of the teenaged daughter
packaging the secret. Alice calls the \texttt{extend} command with a
fresh nonce $n$ in an encrypted session.  She uses the \texttt{create
  key} command constraining that new key to be used only when a
specific value is present in the PCR.  In particular, the constraining
value $cv$ she chooses is the following:
$$ cv = \#(\cn{\mathtt{``obtain"}},\#(n,\fn{val})) $$
where $\fn{val}$ was the PCR value prior the extend command.  She then
encrypts her secret $v$ with this newly created key.

Using typical message passing notation, Alice's part of the protocol
might be represented as follows (where $k'$ denotes the key
created in the second line, and where we still ignore the replay protection):\\

\noindent
$$
\begin{array}{c@{{}\to{}}c@{~:~}l}
\mathrm{A} & \mathrm{TPM} & \enc{\mathtt{``extend"},n}{k}\\
\mathrm{A} & \mathrm{TPM} & \mathtt{``create\ key"},
\#(\mathtt{``obtain"},\#(n,val)) \\
\mathrm{TPM} & \mathrm{A} & k'\\
\mathrm{A} & \mathrm{Parent} & \enc{v}{k'}\\
\end{array}
$$
The parent acts as the adversary in this protocol. We assume he can
perform all the normal Dolev-Yao operations such as encrypting and
decrypting messages when he has the relevant key, and interacting with
honest protocol participants. Most importantly, the parent can use the
TPM commands available in any order with any inputs he likes. Thus he
can extend the PCR with the string \texttt{obtain} and use the key to
decrypt the secret.  Alternatively, he can extend the PCR with the
string \texttt{refuse} and then generate a TPM quote as evidence the
secret will never be exposed. The goal of the Envelope Protocol is to
ensure that once Alice has prepared the TPM and encrypted her secret,
the parent should not be able to both decrypt the secret and also
generate a refusal quote, $\enc{\cn{\mathtt{``quote"}},
  \#(\cn{\mathtt{``refuse"}},\#(n,\fn{val})), \enc{v}{k'}}{\fn{aik}}$.

A crucial fact about the PCR role in this protocol is the
injective nature of the hashing, ensuring that for every $x$
\begin{align} \label{eq:injective} \#(\mathtt{``obtain"},\#(n,val))
  \quad \neq \quad \#(\mathtt{``refuse"}, x )
\end{align}

\section{The TPM Model}
\label{sec:tpm}

In this section we introduce our TPM state theory
$\tstate(\leadsto\nobreak)$ focusing on representing the value of
the PCR and how the TPM commands may change it.

%
%

\begin{figure*}[t]
$$\begin{array}{ll@{{}\typ{}}ll}
\mbox{Sorts:}&
\multicolumn{3}{l}{\mbox{\srt{M}, $\top$, $\srt{A}$, $\srt{S}$, $\srt{D}$,
 $\srt{E}$}}\\
\mbox{Subsorts:}&
\multicolumn{3}{l}{\mbox{$\srt{A}<\top$, $\srt{S}<\top$,
    $\srt{D}<\top$, $\srt{E}<\top$}}\\
\mbox{Operations:}&\boot&\srt{M}&\mbox{TPM boot}\\
&\extend&\top\times\srt{M}\to\srt{M}&\mbox{TPM extend}\\
&(\cdot,\cdot)&\top\times\top\to\top& \mbox{Pairing}\\
&\enc{\cdot}{(\cdot)}&\top\times\srt{A}\to\top&\mbox{Asymmetric encryption}\\
&\enc{\cdot}{(\cdot)}&\top\times\srt{S}\to\top&\mbox{Symmetric encryption}\\
&\invk{(\cdot)}&\srt{A}\to\srt{A}& \mbox{Asymmetric key inverse}\\
&\invk{(\cdot)}&\srt{S}\to\srt{S}& \mbox{Symmetric key inverse}\\
&\#&\srt{\top}\to\srt{S}& \mbox{Hashing}\\
&\cn{a}_i,\cn{b}_i&\srt{A}& \mbox{Asymmetric key constants}\\
&\cn{s}_i&\srt{S}& \mbox{Symmetric key constants}\\
&\cn{d}_i&\srt{D}& \mbox{Data constants}\\
&\cn{e}_i&\srt{E}& \mbox{Text constants}\\
&\tg{i}&\top& \mbox{Tag constants}\\
\mbox{Equations:}&\multicolumn{2}{l}{\invk{\cn{a}_i}=\cn{b}_i\quad
\invk{\cn{b}_i}=\cn{a}_i}
&(i\in\nat)\\
&\multicolumn{2}{l}{\all{k\typ\srt{A}}\invk{(\invk{k})}=k}
&\all{k\typ\srt{S}}\invk{k}=k
\end{array}$$
\caption{Crypto Algebra with State Signature}\label{fig:signature}
\end{figure*}

Fig.~\ref{fig:signature} shows the signature of the order-sorted
algebra used in this paper.  Sort~\srt{M} is the sort of TPM machine
states and sort~$\top$ is the top sort of messages.  Messages of
sort~$\srt{A}$ (asymmetric keys), sort~$\srt{S}$ (symmetric keys),
sort~$\srt{D}$ (data), and sort~$\srt{E}$ (text) are called
\emph{atoms}.  Messages are atoms, tag constants, or constructed using
encryption $\enc{\cdot}{(\cdot)}$, hashing $\#(\cdot)$, and pairing
$(\cdot,\cdot)$, where the comma operation is right associative and
parentheses are omitted when the context permits.

The algebra is the initial quotient term algebra over the signature.
It is easy to show that each term $t$ of the algebra is equal to a
unique term $t'$ with no occurrences of the inverse
operation~$\invk{(\cdot)}$; we choose this $t'$ to be the canonical
representative of $t$.

We use the function {\pcr} to coerce TPM states, which are of
sort~$\srt{M}$, to messages, specifically to symmetric keys of
sort~$\srt{S}$:
\ifllncs
\[
\pcr(\boot) = \cn{s}_0  \qquad
\pcr(\extend(t, m)) =  \#(t,\pcr(m))
\]
\else
$$\begin{array}{r@{{}={}}l}
\multicolumn{2}{c}{\pcr\typ \; \srt{M}\to\srt{S}}\\
\pcr(\boot)&\cn{s}_0\\
\pcr(\extend(t, m))&\#(t,\pcr(m))
\end{array}$$
\fi
where constant~$\cn{s}_0$ is known to all.  Modeling the injectivity
of the hash function (cf.  Equation~\ref{eq:injective}), we postulate
that the function {\pcr} is injective.


The definition of the TPM transition relation~$\tran$ is
\ifieeetrans
\[\begin{array}{llr@{{}={}}lr}
  m_0\tran m_1&\text{iff}&m_1&\boot & \mbox{(\texttt{boot})} \\
&\text{or}&\some{t\typ\top} m_1& \cn{ex}(t,m_0) &
\mbox{(\texttt{extend})} \\
&\text{or}& m_0 &
\multicolumn{2}{@{}l}{m_1\qquad
\mbox{(\texttt{quote}, \texttt{decrypt})}}
\end{array}\]
\else
\[\begin{array}{llr@{{}={}}lr}
  m_0\tran m_1\quad
&\text{iff}&m_1&\boot & \mbox{(\texttt{boot})} \\
&\text{or}&\some{t\typ\top} m_1& \cn{ex}(t,m_0) &
\mbox{(\texttt{extend})} \\
&\text{or}& m_0 &m_1&\quad
\mbox{(\texttt{quote}, \texttt{decrypt})}
\end{array}\]
\fi
The \texttt{create key} command does not interact with the state.

%
%
\ifllncs\begin{sloppypar}\fi
In this framework we prove a crucial property of all executions which
we express in terms of the notion of a state \emph{having} a
message. A state \emph{has} a message if an extend operation with it
is part of the state.
For example, $\cn{ex}(\mathtt{``obtain"},\cn{ex}(v,\cn{bt}))$ has
``obtain'' and $v$, but it does not have ``refuse''.
\ifllncs\end{sloppypar}\fi

An infinite sequence of states~$\pth$ is a \emph{path} if
$\pth(0)=\boot$ and $\all{i\in\nat}(\pth(i),
\ifieeetrans\else\break\fi\pth(i+1))\in{\tran}$.
Paths in this TPM model have several useful properties.  For example,
if a previous state is not a subterm of a state, there must have been
an intervening boot.  Also, if a state has a message, and a previous
state is a boot state, there must have been an intervening transition
that extends with the message.  These two properties can be combined
into the property used by the proof of the Envelope Protocol security
goal: if a previous state is not a subterm of a state that has a
message, there must have been an intervening transition that extends
with the message. Lemma~\ref{lem:prefix boot extend} formalizes this
property in our state theory $\tstate(\leadsto)$, and we proved it
using {\pvs}.


%
%

\begin{lem}[Prefix Boot Extend]\label{lem:prefix boot extend}
$$\begin{array}{l}
\all{\pth\in\fn{path},t\typ\top,i,k\in\nat}
i\leq k\land\mbox{$\pth(k)$ has $t$}\\
\quad\supset\fn{subterm}({\pth(i), \pth(k)})\\
\qquad{}\lor\some{j\in\nat}
i\leq j<k\land\pth(j+1)=\extend(t,\pth(j))
\end{array}$$
\end{lem}

\section{Strand Spaces}
\label{sec:strand spaces}

This section introduces our strand space theory of the envelope
protocol, $\tbnd(\Pi)$. In strand space
theory~\cite{ThayerHerzogGuttman99}, a strand represents the local
behavior of a principal in a single session.  The \emph{trace} of a
strand is a linearly ordered sequence of events
$e_0\Rightarrow\cdots\Rightarrow e_{n-1}$, and an \emph{event} is
either a message transmission $\outbnd t$ or a reception $\inbnd t$,
where~$t$ has sort~$\top$.  A \emph{strand space}~$\Theta$ is a map
from a set of strands to a set of traces.  In the {\pvs} theory of
strand spaces, the set of strands is a prefix of the natural numbers,
so a strand space is a finite sequence of traces.

In a strand space, a node identifies an event.  The \emph{nodes} of
strand space $\Theta$ are $\{(s,i)\mid s\in\sdom(\Theta), 0\leq i <
|\Theta(s)|\}$, and the event at a node is
$\evt_\Theta(s,i)=\Theta(s)(i)$.

A message~$t_0$ is \emph{carried by}~$t_1$, written $t_0\sqsubseteq
t_1$ if~$t_0$ can be extracted from a reception of~$t_1$, assuming the
necessary keys are available.  In other words,~$\sqsubseteq$ is the
smallest reflexive, transitive relation such that $t_0\sqsubseteq
t_0$, $t_0\sqsubseteq (t_0, t_1)$, $t_1\sqsubseteq (t_0, t_1)$, and
$t_0\sqsubseteq\enc{t_0}{t_1}$.
A message \emph{originates} in trace~$c$ at index~$i$ if it is carried
by $c(i)$, $c(i)$ is a transmission, and it is not carried by any event
earlier in the trace.  A message~$t$ is \emph{non-originating} in a
strand space~$\Theta$, written $\fn{non}(\Theta,t)$, if it originates
on no strand.  A message~$t$ \emph{uniquely originates} in a strand
space~$\Theta$ at node~$n$, written $\fn{uniq}(\Theta,t,n)$, if it
originates in the trace of exactly one strand~$s$ at index~$i$, and
$n=(s,i)$.

The model of execution is a bundle.  The pair $\Upsilon=(\Theta,\to)$
is a \emph{bundle} if it defines a finite directed acyclic graph,
where the vertices are the nodes of $\Theta$, and an edge represents
communication~($\rightarrow$) or strand succession~($\Rightarrow$)
in~$\Theta$.  For communication, if $n_0\rightarrow n_1$, then there
is a message~$t$ such that~$\evt_\Theta(n_0)=\outbnd t$
and~$\evt_\Theta(n_1)=\inbnd t$.  For each reception node~$n_1$, there
is a unique transmission node~$n_0$ with $n_0\rightarrow n_1$. %
We use $\prec$ to denote the causal ordering of nodes in a bundle:
the transitive closure of $ \rightarrow \cup \Rightarrow$.
The strand space associated  with a bundle $\Upsilon$ will be denoted
$\Theta_\Upsilon$ unless the association is clear from the context.

%
%

When a bundle is a run of a protocol, the behavior of each strand is
constrained by a role.  Adversarial strands are constrained by roles
as are non-adversarial strands.  A \emph{protocol} is a set of roles,
and a \emph{role} is a set of traces.  A trace~$c$ is an
\emph{instance} of role~$r$ if~$c$ is a prefix of some member of~$r$.
More precisely, for protocol~$P$, we say that bundle
$\Upsilon=(\Theta,\to)$ is a \emph{run of protocol}~$P$ if there
exists a role assignment $\ra\in \sdom(\Theta)\to P$ such that for all
$s\in\sdom(\Theta)$, $\Theta(s)$ is an instance of~$\ra(s)$.
In what follows, we fix the protocol~$P$ and only consider bundles
that are runs of~$P$.

The roles that constrain adversarial behavior are defined by the
functions
\ifllncs
\begin{wrapfigure}{l}{.5\textwidth}\small\vspace{-8mm}
\else
\begin{figure}[t]
\fi
$$\begin{array}{r@{{}={}}l}
\fn{create}(t\typ\base)&\outbnd t\qquad\fn{tag}_i=\outbnd\tg{i}\\
\fn{pair}(t_0\typ\top, t_1\typ\top)&
\inbnd t_0\Rightarrow\inbnd t_1\Rightarrow\outbnd (t_0,t_1)\\
\fn{sep}(t_0\typ\top, t_1\typ\top)&
\inbnd (t_0, t_1)\Rightarrow\outbnd t_0\Rightarrow\outbnd t_1\\
\fn{enc}(t\typ\top, k\typ\srt{A}|\srt{S})&
\inbnd t\Rightarrow\inbnd k\Rightarrow\outbnd \enc{t}{k}\\
\fn{dec}(t\typ\top, k\typ\key)&
\inbnd \enc{t}{k}\Rightarrow\inbnd\invk{k}\Rightarrow\outbnd t\\
\fn{hash}(t\typ\top)&
\inbnd t\Rightarrow\outbnd \#t
\end{array}$$
\caption{Adversary Traces}\label{fig:adversary}
\ifllncs
\vspace{-6mm}
\end{wrapfigure}
\else
\end{figure}
\fi
in Figure~\ref{fig:adversary}.  The adversary can execute all
instances of these patterns.  For the encryption related roles,
$k\typ\key$ asserts that~$k$ is either a symmetric or asymmetric key.
For the create role, $t\typ\base$ asserts that~$t$ is an atom.  Atoms,
characteristically, are what the adversary can create out of thin air
(modulo origination assumptions).

%
%

There is a role for each TPM operation.  We represent them using a
\emph{state-passing style}.  The state-passing style allows {\cpsa} to
do draw conclusions about where states could come from.  Each role
receives a message encoding the state at the time it occurs.  It
transmits a message encoding the state after any state change it
causes.  We do the encoding using a special tag $\tg{0}$ and an
encryption.  For a transition $m_0\tran m_1$, the role contains
$$\cdots\Rightarrow\inbnd\enc{\tg{0},\pcr(m_0)}{\#k}\Rightarrow
\outbnd\enc{\tg{0},\pcr(m_1)}{\#k}\Rightarrow\cdots.$$
Here~$k$ is an uncompromised symmetric key used only in TPM
operations.  The states are encoded as encryptions using the hash
${\#k}$ of $k$.  Tag~$\tg{0}$ is included to ensure that a
state-bearing message is never confused with any other protocol
message.
State-passing style is less restrictive than actual state histories,
since a state-bearing message may be received many times, even if it
is sent only once.

\begin{figure*}[t]
\ifart
\[
\begin{array}{l}
  \fn{boot}(k\typ\srt{S}, p\typ\top)={}\\
  \quad\inbnd\cn{g}_3\Rightarrow\inbnd
  \enc{\tg{0}, p}{\#k}\Rightarrow\outbnd\enc{\tg{0},
    \cn{s}_0}{\#k}\\[1.2ex]

  \fn{extend}(\mathit{sid}\typ\srt{D}, \mathit{tpmk}\typ\srt{A},
  \mathit{esk},k\typ\srt{S},p,t\typ\top)={}\\
  \quad\inbnd (\cn{g}_4,\mathit{tpmk},
  \enc{\mathit{esk}}{\mathit{tpmk}}) \Rightarrow
  \outbnd (\cn{g}_4, \mathit{sid}) \Rightarrow
  \inbnd\enc{\cn{g}_5,t,\mathit{sid}}{\mathit{esk}}\\
  \qquad\Rightarrow\inbnd
  \enc{\tg{0},p}{\#k} \Rightarrow\outbnd\enc{\tg{0},
    \#(t,p)}{\#k}\\[1.2ex]

  \fn{quote}(k\typ\srt{S}, \mathit{aik}\typ\srt{A},
  p,n\typ\top)={}\\
  \quad\inbnd  (\cn{g}_6,n)\Rightarrow\inbnd\enc{\tg{0},p}{\#k}\Rightarrow
  \outbnd\enc{\tg{0},p}{\#k}\Rightarrow \outbnd
  \enc{\cn{g}_6,p,n}{\mathit{aik}}\\[1.2ex]

  \fn{decrypt}(m,t\typ\top, k',\mathit{aik}\typ\srt{A}, k\typ\srt{S})={}\\
  \quad\inbnd (\cn{g}_7,\enc{m}{k'})\Rightarrow\inbnd
  \enc{\cn{g}_8, k', p}{\mathit{aik}}\Rightarrow \inbnd
  \enc{\tg{0},p}{\#k}\Rightarrow \outbnd \enc{\tg{0},p}{\#k}
  \Rightarrow \outbnd m\\[1.2ex]

  \fn{createkey}(k,\mathit{aik}\typ\srt{A}, t\typ\top)={}\\
  \quad\inbnd
  (\cn{g}_9, t)\Rightarrow \outbnd \enc{\cn{g}_8, k,
    t}{\mathit{aik}}
\end{array}
\]
\begin{center}
  \fbox{
    \begin{tabular}{llllllllll}
      $\cn{g}_0$&state&
      $\cn{g}_2$&refuse&
      $\cn{g}_4$&session&
      $\cn{g}_6$&quote&
      $\cn{g}_8$&created\\
      $\cn{g}_1$&obtain&
      $\cn{g}_3$&boot&
      $\cn{g}_5$&extend&
      $\cn{g}_7$&decrypt&
      $\cn{g}_9$&create key
    \end{tabular}
  }
\end{center}
\else
\begin{center}
\begin{minipage}{0.7\textwidth}
$$
\begin{array}{l}
  \fn{boot}(k\typ\srt{S}, p\typ\top)={}\\
  \quad\inbnd\cn{g}_3\Rightarrow\inbnd
  \enc{\tg{0}, p}{\#k}\Rightarrow\outbnd\enc{\tg{0},
    \cn{s}_0}{\#k}\\[1.2ex]

  \fn{extend}(\mathit{sid}\typ\srt{D}, \mathit{tpmk}\typ\srt{A},
  \mathit{esk},k\typ\srt{S},p,t\typ\top)={}\\
  \quad\inbnd (\cn{g}_4,\mathit{tpmk},
  \enc{\mathit{esk}}{\mathit{tpmk}}) \Rightarrow
  \outbnd (\cn{g}_4, \mathit{sid}) \Rightarrow
  \inbnd\enc{\cn{g}_5,t,\mathit{sid}}{\mathit{esk}}\\
  \qquad\Rightarrow\inbnd
  \enc{\tg{0},p}{\#k} \Rightarrow\outbnd\enc{\tg{0},
    \#(t,p)}{\#k}\\[1.2ex]

  \fn{quote}(k\typ\srt{S}, \mathit{aik}\typ\srt{A},
  p,n\typ\top)={}\\
  \quad\inbnd  (\cn{g}_6,n)\Rightarrow\inbnd\enc{\tg{0},p}{\#k}\Rightarrow
  \outbnd\enc{\tg{0},p}{\#k}\Rightarrow \outbnd
  \enc{\cn{g}_6,p,n}{\mathit{aik}}\\[1.2ex]

  \fn{decrypt}(m,t\typ\top, k',\mathit{aik}\typ\srt{A}, k\typ\srt{S})={}\\
  \quad\inbnd (\cn{g}_7,\enc{m}{k'})\Rightarrow\inbnd
  \enc{\cn{g}_8, k', p}{\mathit{aik}}\Rightarrow \\
  \qquad\inbnd
  \enc{\tg{0},p}{\#k}\Rightarrow \outbnd \enc{\tg{0},p}{\#k}
  \Rightarrow \outbnd m\\[1.2ex]

  \fn{createkey}(k,\mathit{aik}\typ\srt{A}, t\typ\top)={}\\
  \quad\inbnd
  (\cn{g}_9, t)\Rightarrow \outbnd \enc{\cn{g}_8, k,
    t}{\mathit{aik}}
\end{array}
$$
\end{minipage}\quad
\begin{minipage}{0.20\textwidth}
  \fbox{\begin{tabular}{ll}
    $\cn{g}_0$&state\\
    $\cn{g}_1$&obtain\\
    $\cn{g}_2$&refuse\\
    $\cn{g}_3$&boot\\
    $\cn{g}_4$&session\\
    $\cn{g}_5$&extend\\
    $\cn{g}_6$&quote\\
    $\cn{g}_7$&decrypt\\
    $\cn{g}_8$&created\\
    $\cn{g}_9$&create key
  \end{tabular}
}
\end{minipage}
\end{center}
\fi
\caption{State-Bearing Traces}\label{fig:state roles}
\end{figure*}

Using these receive-transmit pairs of state-bearing messages the TPM
roles are represented in Fig.~\ref{fig:state roles}, where tag
$\cn{g}_1$ is obtain and tag $\cn{g}_2$ is refuse.  In the \fn{extend}
role, we now show the two initial messages that provide replay
prevention; the TPM supplies a fresh nonce as a session ID that must
appear with the value to be extended into the PCR.  The \fn{createkey}
role does not interact with the state. It simply creates a key that
will be constrained by the state in the boot role.

Alice's role, including the messages to prevent replays, is:
$$
\begin{array}{l}
  \fn{alice}(\mathit{sid},v\typ\srt{D},\mathit{esk}\typ\srt{S},
  k,\mathit{tpmk},\mathit{aik}\typ\srt{A},n\typ\srt{E}
  \ifpriorities, p\typ\top\fi)={}\\
  \quad\outbnd (\cn{g}_4,\mathit{tpmk},
  \enc{\mathit{esk}}{\mathit{tpmk}}) \Rightarrow
  \inbnd (\cn{g}_4, \mathit{sid})\\
  \qquad \Rightarrow \outbnd
  \enc{\cn{g}_5,n,\mathit{sid}}{\mathit{esk}}
  \Rightarrow
  \outbnd(\cn{g}_9, \#(\cn{g}_1,\#(n,\initstate)))\\
  \qquad \Rightarrow
  \inbnd \enc{\cn{g}_8,\#(\cn{g}_1,\#(n,\initstate))}{\mathit{aik}}
  \Rightarrow \outbnd \enc{v}{k}
\end{array}
$$
The parameters $\mathit{sid}$ and~$\mathit{tpmk}$ help prevent
replays.  To make formulas more comprehensible, we omit them.

\section{CPSA}
\label{sec:cpsa} 

\ifllncs
\else
\begin{figure}[t]
    \begin{center}
    \includegraphics{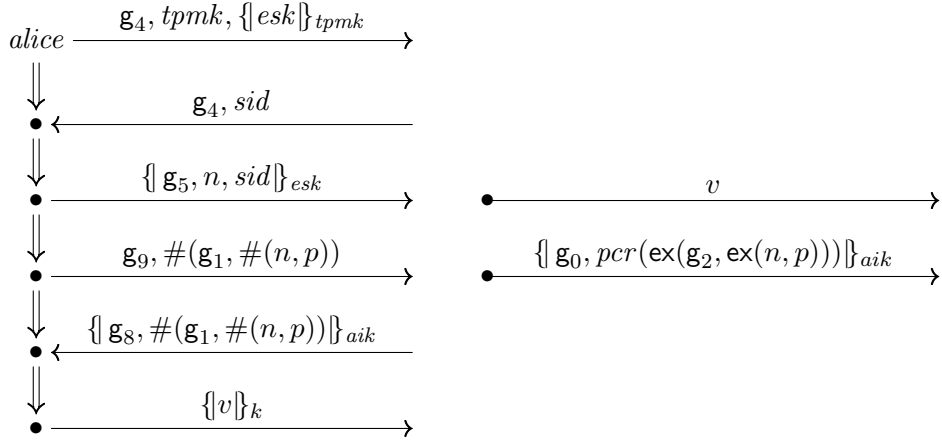}
  \end{center}
  \caption{Alice Point-Of-View}\label{fig:alice pov}
\end{figure}
\fi

This section discusses how we use our analysis tool {\cpsa} to infer
results in the theory $\tbnd(\Pi)$. {\cpsa} carries out enrich-by-need
analysis, and characterizes the set of bundles consistent with a
partial description of a bundle.

These partial descriptions are called \emph{skeletons}. {\cpsa} takes
as input an initial skeleton $\skel_0$, and when it terminates it
outputs a set of more descriptive skeletons $\{\skelB_i\}_{i\in I}$.
They have the property that any bundle containing the structure in the
initial skeleton $\skel_0$ also contains all the structure in one of
the output skeletons $\skelB_i$.  In particular, it infers all of the
non-adversarial behavior that must be present in any bundle satisfying
the initial description. Of course for some initial skeletons
$\skel_0$, there may be no bundles that are consistent with them. In
this case, {\cpsa} outputs the empty set.

The security goal for the Envelope Protocol is that a run of Alice's
role should ensure that the secret and the refusal certificate are not
both available:
\begin{goal}\label{goal:alice}
  Consider the following events:
  \begin{itemize}
    \item An instance of the Alice role runs to completion, with
    secret $v$ and nonce $n$ both freshly chosen;
    \item $v$ is observed unencrypted;
    \item the refusal certificate \ifieeetrans
      $$\enc{\cn{\mathtt{``quote"}},
      \#(\cn{\mathtt{``refuse"}},\#(n,\fn{val})),
      \enc{v}{k'}}{\fn{aik}}$$
      \else
      $\enc{\cn{\mathtt{``quote"}},
      \#(\cn{\mathtt{``refuse"}},\#(n,\fn{val})),
      \enc{v}{k'}}{\fn{aik}}$
      \fi
      is {\ifllncs ob-\break served\else observed\fi} unencrypted.
  \end{itemize}
  These events, which we call jointly $\skel_0$, are not all present
  in any execution.
\end{goal}
We can feed {\cpsa} an input skeleton $\skel_0$ representing this
undesirable situation.
\ifllncs
\else
The skeleton~$\skel_0$ is visualized in Fig.~\ref{fig:alice pov}.
\fi

We would hope {\cpsa} could determine that no bundles are consistent
with this input $\skel_0$ and return the empty set. However, our
technique of using state-bearing messages to represent the TPM state
transitions underconstrains the set of possible state paths.  For this
reason, {\cpsa} actually produces one skeleton in its output.  This
skeleton represents some activity that must have occured within the
TPM in any bundle conforming to the initial skeleton. It contains an
instance of the decrypt role (to explain the secret leaking), an
instance of the quote role (to explain the creation of the refusal
token), and several instances of the extend role (to explain how the
TPM state evolved in order to allow the other two operations).

\begin{figure*}[t]
  \begin{center}
    \includegraphics{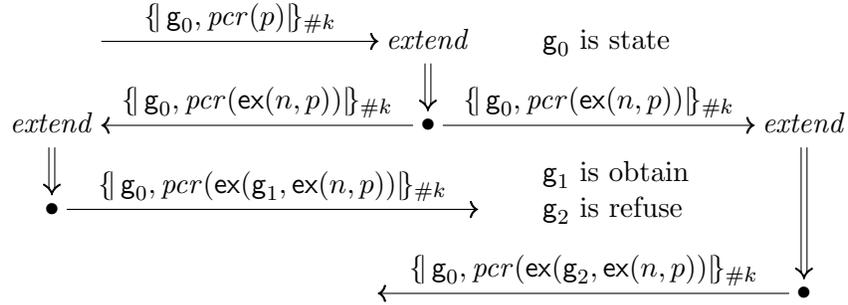}
    \caption{State Splitting}\label{fig:state split}
  \end{center}
\end{figure*}

Fig.~\ref{fig:state split} displays the relevant portion of
{\cpsa}'s output displaying only the state-bearing nodes of the extend
strands inferred by {\cpsa}. Notice that two of the extend strands
branch off from the third strand. This is a state split in which a
single state evolves in two distinct ways. The technique of using
state-bearing messages is not sufficient to preclude this possibility.

{\cpsa}'s enrich-by-need approach is a form of model finding, rather
than theorem proving. In order to use {\cpsa}'s results to our
advantage we need to express its conclusions in the logical theory
$\tbnd(\Pi)$. For that purpose we transform our skeletons into
formulas in order-sorted logic and define what it means for a bundle
to satisfy these formulas. The sorts are the message algebra sorts
augmented with a sort~\srt{Z} for strands and sort~\srt{N} for nodes.
The atomic formula $\cn{htin}(z,h,c)$ asserts that strand~$z$ has a
length of at least~$h$, and its trace is a prefix of trace~$c$. The
formula $n_0\ll n_1$ asserts node~$n_0$ precedes node~$n_1$.  The
formula $\cn{non}(t)$ asserts that message~$t$ is non-originating, and
$\cn{uniq}(t,n)$ asserts that message~$t$ uniquely originates at
node~$n$.  Finally, the formula $\cn{sends}(n,t)$ asserts that the
event at node~$n$ is a transmission of message~$t$.  The roles of the
protocol serve as function symbols. A skeleton $\skel$ is represented
by the conjunction of all facts true in the skeleton.

We encode an entire {\cpsa} analysis by first encoding the input
skeleton $\skel_0$ and the output skeletons $\{\skelB_i\}_{i\in
  I}$. The analysis is then encoded as an implication.  A formula
$\Phi_0$ describing the input $\skel_0$, is the hypothesis of the
conditional.  The disjunction of the formulas $\Psi_i$ describing the
outputs $\{\skelB_i\}_{i\in I}$ form the conclusion.  When {\cpsa}
discovers that there are no bundles compatible with the initial
skeleton, the conclusion is encoded as the empty disjunction, $\bot$.

\begin{figure}[t]
$$\begin{array}{l@{\quad}l}
\Upsilon,\alpha\models x=y&\mbox{iff $\alpha(x)=\alpha(y)$;}\\
\Upsilon,\alpha\models\cn{htin}(z,h,c)
&\mbox{iff $|\Theta_\Upsilon(\alpha(z))|\geq \alpha(h)$ and}\\
&\mbox{\phantom{iff} $\Theta_\Upsilon(\alpha(z))$ is a prefix of $\alpha(c)$;}\\
\Upsilon,\alpha\models n_0\ll n_1
&\mbox{iff $\alpha(n_0)\prec_\Upsilon\alpha(n_1)$;}\\
\Upsilon,\alpha\models\cn{non}(t)
&\mbox{iff $\fn{non}(\Theta_\Upsilon,\alpha(t))$;}\\
\Upsilon,\alpha\models\cn{uniq}(t,n)
&\mbox{iff $\fn{uniq}(\Theta_\Upsilon,\alpha(t),\alpha(n))$;}\\
\Upsilon,\alpha\models\cn{sends}(n,t)
&\mbox{iff $\evt_{\Theta_\Upsilon}(\alpha(n))=\outbnd\alpha(t)$.}
\end{array}$$
\caption{Satisfaction}\label{fig:satisfaction}
\end{figure}

The satisfaction relation is defined using the clauses in
Fig.~\ref{fig:satisfaction}. It relates a bundle, a variable
assignment, and a formula: $\Upsilon,\alpha\models\Phi$.
%
A bundle~$\Upsilon$ is described by a skeleton iff the skeleton's
sentence~$\Phi$ satisfies~$\Upsilon$, written $\Upsilon\models\Phi$.
%

The formula $\Phi_0$ that specifies the initial skeleton relevant to
the Envelope Protocol security goal
is
\begin{equation}
  \label{eq:phi:0}
  \begin{array}{l}
    \cn{htin}(z,4,\fn{alice}(v,\mathit{esk},k,\mathit{aik},n
    \ifpriorities,\initstate\fi))\land\cn{sends}(n_1,v)\\
    \quad\land \cn{sends}(n_2,\enc{\tg{0},\fn{pcr}(\extend(\tg{2},\extend(n,\initstate)))}{\mathit{aik}})\\
    \quad\land\cn{non}(\mathit{aik})\land\cn{non}(\mathit{esk})\\
    \quad\land\cn{uniq}(n,(z,1))\land\cn{uniq}(v,(z,4)),
  \end{array}
\end{equation}
where
${v\typ\srt{D},\mathit{esk}\typ\srt{S},k,\mathit{aik}\typ\srt{A},n\typ\srt{E},
  \ifpriorities p\typ\top,\fi z\typ\srt{Z},n_1,n_2\typ\srt{N}}$.

The output skeleton $\skelB_1$ is much larger and its formula $\Psi_1$
is correspondingly large. The relevant part of this formula
representing the fragment in Fig.~\ref{fig:state split} is
\begin{equation}
  \label{eq:psi:1}
\begin{array}{l}
  \cn{htin}(z_1,3,\fn{extend}(\mathit{esk},k,\fn{pcr}(\initstate),n))\\
  \quad\land\cn{htin}(z_2,3,\fn{extend}(\mathit{esk},k,\fn{pcr}(\extend(n,\initstate)),\tg{1}))\\
  \quad\land\cn{htin}(z_3,3,\fn{extend}(\mathit{esk},k,\fn{pcr}(\extend(n,\initstate)),\tg{2})),
\end{array}
\end{equation}
where ${\mathit{esk},k\typ\srt{S}, \ifpriorities
  p\typ\top,\fi n\typ\srt{E}, z_1,z_2,z_3\typ\srt{Z}}$.
The full formula for $\skelB_1$ has more conjuncts.

\ifart\begin{sloppypar}\fi
Let the vector $\overline{x}$ contain the variables that appear free
in $\Phi_0$, and possibly also in $\Psi_1$, and let the vector
$\overline{y}$ contain the variables that occur free in $\Psi_1$ only.
Summarizing {\cpsa}'s analysis for the Envelope Protocol in
$\tbnd(\Pi)$, we have:
\begin{lem}\label{lem:alice shape}
  $\all {\,\overline{x}}(\Phi_0 \,\supset\,
  \some{\,\overline{y}}\Psi_1)$, where $\Phi_0,\Psi_1$ are as in
  formulas~\ref{eq:phi:0}--\ref{eq:psi:1}.
\end{lem}
However, unlike Lemma~\ref{lem:prefix boot extend}, this lemma was not
derived within {\pvs}.  Rather, it is true if {\cpsa}'s analysis is
correct.  We import it into {\pvs} as an axiom.
\ifart\end{sloppypar}\fi

Lemma~\ref{lem:alice shape} is however something capable of direct
proof within {\pvs} as a theorem of $\tbnd(\Pi)$.  Indeed, there is
precedent for constructing proofs of this sort.  Meier et
al.~\cite{meier2013efficient} show how to instrument a different
protocol analysis tool, called Scyther~\cite{cremers2012operational},
so that each step it takes generates a lemma in the Isabelle proof
system.  Then, they use reusable results proved once within Isabelle
to discharge these lemmas.  Curiously, one of the main lemmas, the
authentication test theorem in an earlier form, has already been
established within {\pvs}~\cite{jacobs2009semantics}.  Thus, it
appears possible, although a substantial undertaking, to transform
{\cpsa} from a central piece of our analysis infrastructure to a
heuristic to guide derivations within {\pvs}.

\section{Reasoning About Messages and State}
\label{sec:pvs}

This section presents some details of the theory
$\tannot(\Pi,\tran)$. We then show how the previous
lemmas combine allowing us to conclude that the security goal of
the Envelope Protcol is achieved.

In $\tannot(\Pi,\tran)$, the state transitions associated with a
protocol are specified by annotating some events in a role of $\Pi$
with a subset of the transition relation~$\tran$.  The reason for
annotating events with a subset of the transition relation, rather
than an element, will be explained at the end of this section.  We
use~$\bot$ for an event that is not annotated, and $\up a$ for an
event that is annotated with~$a$.  The events that are annotated are
the transmissions associated with receive-transmit pairs of
state-bearing messages.
$$\begin{array}{*{7}{c@{}}}
\cdots&{}\Rightarrow{}&\inbnd\enc{\tg{0},\pcr(m_0)}{\#k}&{}\Rightarrow{}&
\outbnd\enc{\tg{0},\pcr(m_1)}{\#k}&{}\Rightarrow{}&\cdots\\
\bot&&\bot&&\up\{(m_0,m_1)\}\cap{\tran}&&\bot
\end{array}$$

\ifieeetrans
\begin{figure*}[t]
  \begin{center}
    \includegraphics{envelope\grphext-1.mps}
  \end{center}
\caption{Inferred Extend Strand}\label{fig:inferred strand}
\end{figure*}
\fi

A node in a bundle inherits its annotation from its role.  The set of
nodes in~$\Upsilon$ that are annotated is $\fn{anode}(\Upsilon)$, and
$\fn{anno}(\Upsilon,n,a)$ asserts that node~$n$ in~$\Upsilon$ is
annotated with some $a\subseteq\tran$.  In the Envelope Protocol, a
node annotated by a TPM extend role cannot be an instance of any other
role.

Our goal is to reason only with bundles that respect state semantics.
A bundle $\Upsilon$ with a transition annotating role assignment is
\index{compatible bundle}\emph{compatible}~\cite[Def.~11]{Guttman12}
with transition relation~$\tran$ if there exists $\ell\in\nat$,
$f\in\fn{anode}(\Upsilon)\to\{0,1,\ldots,\ell-1\}$, and
$\pi\in\fn{path}$ such that
\begin{enumerate}
\item $f$ is bijective;
\item $\all{n_0,n_1\in\fn{anode}(\Upsilon)}
n_0\prec n_1\Longleftrightarrow f(n_0)<f(n_1)$;
\item $\all{n\in\fn{anode}(\Upsilon),a\subseteq{\tran}}$\\
$\mbox{}\quad\fn{anno}(\Upsilon,n,a)\supset
(\pi(f(n)),\pi(f(n)+1))\in a$.
\end{enumerate}
A bundle that satisfies $\tannot(\Pi,\tran)$ is a compatible
bundle.


Because the function~$f$ is bijective, all annotated nodes in a
compatible bundle are totally ordered.  Looking back at
Fig.~\ref{fig:state split}, either the nodes in the leftmost
strand precede the nodes in the rightmost strand or succeed them.



The compatible bundle assumption allows one to infer the existence of
nodes that are not revealed by {\cpsa}.  In the case of the Envelope
Protocol this is done by importing the Prefix Boot Extend Lemma
(Lemma~\ref{lem:prefix boot extend}) from $\tstate(\tran)$ into the
strand space world by proving the following lemma (stated here in
plain English) within $\tannot(\Pi,\tran)$ using {\pvs}.  Its proof
uses the full content of compatibility.

\begin{lem}[Bridge, informally]\label{lem:bridge}
  Let $\Upsilon$ be a compatible bundle, containing two annotated
  nodes, $n_0\prec n_1$, where $n_1\!$'s state has a value $t$.  Then
  either $n_0\!$'s state is a subterm of $n_1\!$'s state, or else
  there is an \emph{extend} node between them that incorporates~$t$.
 \end{lem}
 %
%


 This Bridge Lemma implies there is another extend strand
 between the two strands that represent the state split. This theorem
 is also proved with {\pvs} in $\tannot(\Pi,\tran)$; however,
 syntactically it is a sentence of the language of $\tbnd(\Pi)$.  That
 is, $\tannot(\Pi,\tran)$ adds information to $\tbnd(\Pi)$,
 because $\tannot(\Pi,\tran)$'s models are only the compatible
 bundles.  The theorem is the following.


%
%
\begin{thm}[Inferred Extend Strand]\label{thm:inferred strand}
\begingroup\rm
$$\begin{array}{l}
\all{z_0,z_1\typ\srt{Z},t,t_0,t_1\typ
  \top,m_0,m_1\typ\srt{M},\mathit{esk}_0,\mathit{esk}_1,k_0,k_1\typ\srt{A}}\\
\quad\cn{htin}(z_0,2,\fn{extend}(\mathit{esk}_0,k_0,\pcr(m_0),t_0))\\
\qquad{}\land\cn{htin}(z_1,2,\fn{extend}(\mathit{esk}_1,k_1,\pcr(m_1),t_1))\\
\qquad{}\land(z_0,1)\ll(z_1,0)\land\mbox{$m_1$ has $t$}\\
\qquad{}\supset\fn{subterm}(\extend(t_0,m_0),m_1)\\
\qquad\quad{}\lor\some{z\typ\srt{Z},m\typ\srt{M},\mathit{esk},k\typ\srt{A}}\\
\qquad\qquad\cn{htin}(z,2,\fn{extend}(\mathit{esk},k,\pcr(m),t))\\
\qquad\qquad\quad{}\land(z_0,1)\ll(z,0)\land(z,1)\ll(z_1,0)
\end{array}$$
\endgroup
\end{thm}

\ifieeetrans
\else
\begin{figure*}[t]
  \begin{center}
    \includegraphics{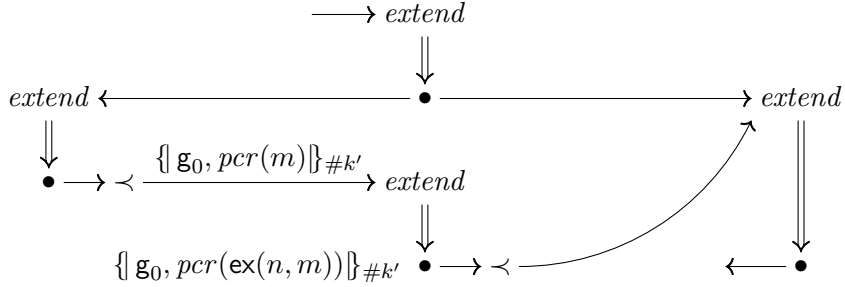}
  \end{center}
\caption{Inferred Extend Strand}\label{fig:inferred strand}
\end{figure*}
\fi

Theorem~\ref{thm:inferred strand} implies that Fig.~\ref{fig:state
  split} has an additional \emph{extend} strand, as shown in
Fig.~\ref{fig:inferred strand}.  Restarting {\cpsa} with $\skel_0$
enriched with all of this additional information, we learn that no
such execution is possible.  This justifies Security
Goal~\ref{goal:alice}.

\para{Our Method.}  We have now completed an illustration of the
hybrid method for analyzing a protocol with state.  We took the
following key steps.
\begin{enumerate}
  \item We defined states and a transition relation representing a
  {TPM} fragment.  We proved a key lemma (Lemma~\ref{lem:prefix boot
    extend}) in the resulting theory $\tstate(\tran)$.
  \item We defined the envelope protocol as a {\pvs} theory
  $\tbnd(\Pi)$.  We encoded the states as certain encrypted messages,
  and used state-passing to represent the actions of the TPM in
  protocol roles.  The encoding function is an injective function $g$.
  We connect $\cdots -t_0\Rightarrow +t_1\cdots$, as a state-passing
  representation, with $\tstate(\tran)$ by annotating the role with
  the annotation:\label{enum:annotate}
  $$\{(m_0,m_1)\mid t_0=g(m_0)\land t_1=g(m_1)\}\cap{\tran}.$$
  We prove bridge lemmas along the lines of Lemma~\ref{lem:bridge}.
  \item Independently, we define $\Pi$ in the {\cpsa} input language,
  and query {\cpsa} with a starting point $\skel_0$ as in our security
  goal.  We translate the results in the form of state analysis
  sentences such as Lemma~\ref{lem:alice shape}, which we use within
  {\pvs} as axioms.
  \item From a state analysis sentence and bridge lemmas, we deduce
  conclusions about all compatible bundles of $\Pi$ and $\tran$.
  Thm.~\ref{thm:inferred strand} was an example.  These theorems may
  already establish our security goals.
  \item Alternatively, the conclusions about compatible bundles may
  give us an enriched starting point, which we can bring back into
  {\cpsa}, as we did here to determine that Security
  Goal~\ref{goal:alice} is achieved, and $\skel_0$ cannot appear in
  any compatible bundle.
\end{enumerate}
We have also applied this method to several simple protocols besides
the Envelope Protocol.  The steps in applying the method are always
the same.  While the application of these ideas is routine, it is
quite time consuming.  A goal of future research is to automate much
more of the method.

But why annotate events with subsets of the transition relation rather
than elements of it?  The \emph{extend} role does not guarantee it
receives a state-bearing message of the form
$\enc{\tg{0},\pcr(m_0)}{\#k}$.  It says only that the incoming message
has the form $\enc{\tg{0},t_0}{\#k}$.  We must eliminate strands in
which~$t_0$ is not in the range of the {\pcr} function.  That is why
we use the annotation shown in Step~\ref{enum:annotate}.

A bundle in which a received state encoding message is not in the
range of the {\pcr} function will have a node annotated with the empty
set.  This bundle does not respect state semantics and is eliminated
from consideration by the definition of compatibility.

%
%



\section{Related Work and Conclusion}
\label{sec:conclusion}

\para{Related Work.}\label{sec:related_work}
The problem of reasoning about protocols and state has been an
increasing focus over the past several years.  Protocols using Trusted
Platform Modules (TPMs) and other hardware security modules (HSMs)
have provided one of the main motivations for this line of work.

A line of work was motivated by HSMs used in the banking
industry~\cite{herzog2006applying,youn2007robbing}.  This work
identified the effects of persistent storage as complicating the
security analysis of the devices.  Much work explored the significance
of this problem in the case of PKCS \#11 style devices for key
management~\cite{cortier2007automatic,cortier2009generic,froschle2011reasoning}.
These papers, while very informative, exploited specific
characteristics of the HSM problem; in particular, the most important
mutable state concerns the \emph{attributes} that determine the usage
permitted for keys.  These attributes should usually be handled in a
monotonic way, so that once an attribute has been set, it will not be
removed.  This justifies using abstractions that are more typical
of standard protocol analysis.

In the TPM-oriented line of work, an early example using an
automata-based model was by G{\"u}rgens et
al.~\cite{gurgens2007security}.  It identified some protocol failures
due to the weak binding between a TPM-resident key and an individual
person.  Datta et al.'s ``A Logic of Secure
Systems''~\cite{datta2009logic} presents a dynamic logic in the style
of PCL~\cite{DattaEtAl05} that can be used to reason about programs
that both manipulate memory and also transmit and receive
cryptographically constructed messages.  Because it has a very
detailed model of execution, it appears to require a level of effort
similar to (multithreaded) program verification, unlike the less
demanding forms of protocol analysis.

M{\"o}dersheim's set-membership
abstraction~\cite{modersheim2010abstraction} works by identifying all
data values (e.g.~keys) that have the same properties; a change in
properties for a given key $K$ is represented by translating all facts
true for $K$'s old abstraction into new facts true of $K$'s new
abstraction.  The reasoning is still based on monotonic methods
(namely Horn clauses).  %
Thus, it seems not to be a strategy for reasoning about TPM usage, for
instance in the envelope protocol.

The paper~\cite{Guttman12} by one of us developed a theory for
protocols (within strand spaces) as constrained by state transitions,
and applied that theory to a fair exchange protocol.  It introduced
the key notion of \emph{compatibility} between a protocol execution
(``bundle'') and a state history.  In the current paper we will also
rely on the same notion of compatibility, which was somewhat hidden
in~\cite{Guttman12}.  However, the current paper does not separate the
protocol behavior from state history as sharply as
did~\cite{Guttman12}.

A group of papers by Ryan with Delaune, Kremer, and
Steel~\cite{delaune2011formal,DelauneEtAl2011}, and with Arapinis and
Ritter~\cite{arapinis2011statverif} aim broadly to adapt ProVerif for
protocols that interact with long-term state.
ProVerif~\cite{Blanchet01,AbadiBlanchet05} is a Horn-clause based
protocol analyzer with a monotonic method: in its normal mode of
usage, it tracks the messages that the adversary can obtain, and
assumes that these will always remain available.  Ryan et al.~address
the inherent non-monotonicity of adversary's capabilities by using a
two-place predicate $\att(u,m)$ meaning that the adversary may possess
$m$ at some time when the long-term state is $u$.
In~\cite{arapinis2011statverif}, the authors provide a compiler from a
process algebra with state-manipulating operators to sets of Horn
clauses using this primitive.  In~\cite{DelauneEtAl2011}, the authors
analyze protocols with specific syntactic properties that help ensure
termination of the analysis.  In particular, they bound the state
values that may be stored in the TPMs.  In this way, the authors
verify two protocols using the TPM, including the envelope protocol.

%
%

One advantage of the current approach relative to the
ProVerif approach is that it works within a single
comprehensive framework, namely that of strand spaces.  Proofs about
state within {\pvs} succeeded only when definitions and lemmas were
properly refined, and all essential details represented.  As a result,
our confidence is high that our proofs about protocols have their
intended meaning.

%
%

%
%


\para{Conclusion.}  The proof of the Envelope Protocol security
goal presented here shows a detailed example of our method for
applying {\cpsa} to systems that include a state component.  {\cpsa}
was coupled with about 2400 lines of {\pvs} specifications to produce
a proof of a difficult security goal.  The method is sound due to the
use of the common foundation of strand space theory for all reasoning.

%
%
The approach could be improved in two main ways.  First, the proofs
within {\pvs} are strenuous.  We would like to develop a method in
which---apart perhaps from a few key reusable lemmas in the state
theory $\tstate(\leadsto)$---the remainder of the reasoning concerning
both state and protocol behavior occurs automatically in {\cpsa}'s
automated, enrich-by-need manner.  Second, there is some artificiality
in the state-threading representation that we have used here.  It
requires the protocol description to make explicit the details of the
full state, and to express each state change in a syntactic,
template-based form.  Moreover, the state information is also
redundantly encoded in the annotations that appear in
$\tannot(\Pi,\leadsto)$.  Our earlier work~\cite{Guttman12} instead
encapsulated all of the state information in a labeled transition
relation.  The protocol definitions contain only a type of ``neutral
node'' which are neither transmissions nor receptions.  These nodes
are associated with the same labels as appear in labeled transitions.
This allows us to define ``compatibility,'' and to work with protocol
and state definitions as independent modules.  We intend also to
explore this style of definition.

\para{Acknowledgment.}  We are grateful to Ed Zieglar for discussions
and support.

\bibliography{secureprotocols}
\bibliographystyle{plain}



\end{document}